# PREVENTING OTHERS FROM COMMERCIALIZING YOUR INNOVATION: EVIDENCE FROM CREATIVE COMMONS LICENSES

**Erdem Dogukan Yilmaz**

Erasmus University Rotterdam, RSM

**Tim Meyer**

University of St.Gallen

**Milan Miric**

USC, Marshall

### Abstract

Online innovation communities are an important source of innovation for many organizations. While contributions to such communities are typically made without financial compensation, these contributions are often governed by licenses such as *Creative Commons* that may prevent others from building upon and commercializing them. While this can diminish the usefulness of contributions, there is limited work analyzing what leads individuals to impose restrictions on the use of their work. In this paper, we examine innovators imposing restrictive licenses within the 3D-printable design community *Thingiverse*. Our analyses suggest that innovators are more likely to restrict commercialization of their contributions as their reputation increases and when reusing contributions created by others. These findings contribute to innovation communities and the growing literature on property rights in digital markets.

# 1. INTRODUCTION

Online innovation communities where individuals create and share contributions are an important source of innovation for many organizations (Altman, Nagle and Tushman, 2022; Gambardella, Raasch and von Hippel, 2016; Jeppesen and Frederiksen, 2006; Murray and O'Mahony, 2006, Nagle, 2018). Recent advances in digitization, and the associated decrease in the cost of design and communication (Baldwin and Von Hippel 2011), have made these communities become even more prevalent across a variety of different contexts that go beyond purely digital settings such as software development (Lakhani, 2016).

A key feature of these communities is that individuals typically create their contributions and make them available within the community without any financial compensation. However, sharing their contributions with a community for free does not necessarily mean that all intellectual property rights are given up by the original innovators[1]. Contributions to innovation communities are often governed by open-source licenses, such as *Creative Commons*, which specify whether and how the contribution may be used (i.e., utilized in its original form), reused (i.e., utilized as a building block for other contributions that build on it) or commercialized (i.e., utilized for commercial purposes, either in its original form or as a building block) by others. For example, when photographers share images on *Flickr*, they choose among different licenses that state whether others are allowed to use, reuse or commercialize these creations. Similarly, when contributing software to *GitHub*, software developers specify how their code may be used, reused or commercialized.

Although many individuals freely release contributions to online communities, oftentimes they explicitly choose non-commercial licenses, which allow others to freely use and reuse their contributions, but restrict them from commercializing these contributions. This can pose a significant challenge for firms, as their well-documented potential to gain commercial value from community contributions (Fosfuri, Giarratana and Luzzi, 2008; Nagle, 2018) is significantly reduced if such restrictive non-commercial licenses are imposed on contributions (Lerner and Tirole, 2002). In order to benefit from innovation communities, it is therefore important for organizations to understand when and why non-commercial restrictions are imposed on community contributions. In this paper, we study the factors that shape the decisions of individuals in innovation communities to use non-commercial licenses.

---

[1] As our paper focuses on the actions of individuals in innovation communities, we use the terms "individual" and "innovator" interchangeably. We acknowledge that there can be meaningful distinctions between the two terms in other settings.

Open-source licenses such as *Creative Commons* are an important form of property rights (Contreras, 2022; Lessig, 2004; Stallman, 2002), and are often thought to be critical to how these communities function. This is because they incentivize individuals to contribute, knowing that they have a right to attribution (credit), as well as discretion for how their contributions are used, reused or even commercialized. This is particularly important as much innovation relies on cumulativeness and reuse of existing innovation (Baldwin and Von Hippel, 2011; Murray and O'Mahony, 2007). There are cases where innovations from online communities have been reused and developed into commercial products across industries, from medical devices, scientific instruments, semiconductors, software and sports equipment (Von Hippel, 2005). At the same time, for contributions released to innovation communities to become commercial products which are marketed and sold, it is necessary that the original contributors do not restrict commercialization outside the community. Understanding the factors that lead individuals to impose non-commercial restrictions is therefore an important issue for online communities and the innovation process more broadly.

Previous research on innovation communities has characterized the motivation of individuals to contribute as a tradeoff between the costs and benefits of creating contributions and releasing them to the community (Lerner and Tirole, 2002). Such benefits may be purely intrinsic (e.g., the enjoyment derived from solving a challenging problem or identifying yourself with open-source ideology) or more extrinsic (e.g., indirect monetary benefits such as reputation or career opportunities (Belenzon and

Schankerman, 2015; Wasko and Faraj, 2005; Xu, Nian and Cabral, 2020)). While existing studies have focused primarily on the factors that motivate individuals to contribute to these communities in the first place (Goes, Guo and Lin, 2016; Lakhani and von Hippel, 2004; Roberts, Hann and Slaughter, 2006; Zhang and Zhu, 2011), there has been less work looking at the types of licenses that individuals impose on their contributions. Studies examining licensing choices in online communities have predominantly focused on project-level licenses and the motivations of individuals to participate in projects based on the associated license (Belenzon and Schankerman, 2015; Sen, Subramaniam and Nelson, 2008). Research however has not studied what drives innovators to choose specific licenses for a contribution *after* it has been created, and whether they allow others to commercialize these contributions.

Online innovation communities are dependent on innovators creating (high-quality) contributions and building on each other's work. Previous research has highlighted that both of these processes are often driven by mechanisms that generate extrinsic motivations for contributors within the community. For instance, individuals might be driven to contribute more or better-quality creations as they expect this to increase their reputation within the community (Lakhani and von Hippel, 2004; Lakhani and Wolf, 2005). Similarly, the ability of individuals to build on each other's work is related to community-based expectations of reciprocity and the idea of exchanging one's work with others (Faraj and Johnson, 2011; Ren, Kraut and Kiesler, 2007). While these mechanisms can foster activity *within* the community and ultimately have a positive effect on the overall quality of contributions, it remains unclear how these two mechanisms affect the licensing choices of creators and thus the ability of other parties to derive commercial value from these contributions *outside* the community.

Therefore, in this paper, we focus on precisely these two mechanisms and study how innovator reputation and the question whether she builds on (i.e., reuses) others' contributions, shape the decision to impose non-commercial licenses. Specifically, we argue that when individuals have limited reputation within the community, they can benefit from others reusing or even commercializing their contributions, as this can increase their visibility and may ultimately lead to reputational gains. These individuals will therefore lean more towards allowing commercialization. On the other hand, individuals with an established reputation in the community do not have the same need to trade off their rights for additional reputation. These individuals lean more towards imposing non-commercial licenses, as their contributions tend to receive more attention and have potentially higher commercial value, which in turn heightens the potential cost for these users if their contributions are commercialized by others. In terms of contribution origin, we differentiate between contributions that are created from scratch and contributions that reuse (i.e., build upon) contributions of others[2]. When individuals reuse contributions of others, the resulting contributions are typically more complete and more polished, which might make the creators more worried about others commercializing them and makes it more likely that they impose non-commercial licenses. On the other hand, individuals who develop a contribution from scratch typically face fewer concerns about commercialization due to the unique and unrefined nature of their creations. As such, they may be less likely to impose non-commercial licenses, thereby aligning more with expectations of openness and exchange within the community.

To test these predictions, we use data from individual contributions to the three-dimensional (3D) printable design community, *Thingiverse*, that is run by one of the largest producers of 3D printers, *Makerbot Industries*. Individual contributors have freedom to choose among different types of *Creative Commons* licenses, including licenses that restrict the ability of others to commercialize these contributions ("Non-Commercial" clauses in *Creative Commons* licenses). We study a sample of 182,453 designs shared within the community, created by 30,093 individual contributors. We find that, as individuals increase in reputation, they become less likely to allow others to commercialize their contributions. Similarly, when individuals have created their contribution based on the contributions of others, they also become less likely to allow others to commercialize their contribution. These results are robust to a variety of specifications including fixed effects or correlated random effects models. Interestingly, we find that individual level fixed effects explain much of the decision to allow others to commercialize their contributions, consistent with the idea that this is driven by individual motivations or preferences. As an additional check, we use an instrumental variable approach that exploits the unexpected featuring of some designs on the main page of the *Thingiverse* website, which leads to a sudden increase in reputation for the designer but does not directly affect the licensing choice. Lastly, we perform a number of robustness checks and analyses for other types of licenses.

_______________

[2] Contributions that are created by building upon other contributions are often referred to as "derivative contributions" or simply "derivatives".

This paper makes several contributions. First, this paper contributes to the literature on innovation communities. Many previous studies have focused on the static factors such as ex-ante motivation to participate in the community in explaining individual contribution (Gambardella et al., 2016; Lakhani and Wolf, 2005; Roberts et al., 2006). This paper extends these findings in two important ways. First, this paper is among the few recent studies that have considered the more dynamic features that shape individual contributions as the community advances (Miric and Jeppesen, 2023; Nagaraj and Piezunka, 2020). Our findings are consistent with the idea that, as extrinsic motivations driven by reputation mechanisms in user communities become more prominent, this may impair intrinsic motivations (Bénabou and Tirole, 2003; Deci and Ryan, 1985; Jeppesen and Frederiksen, 2006). Moreover, our findings suggest that reputation-based incentive mechanisms (such as likes or ranks) may push individuals to contribute (Lerner & Tirole, 2002; Roberts et al., 2006) but can, in turn, also lead individuals to restrict commercialization of such contributions. Second, many existing studies consider the decision to contribute, but do not consider the restrictions that the creator may impose on their contributions. Previous studies on intellectual property rights in such communities have mostly focused on how project licenses affect individuals' motivation to contribute (Belenzon and Schankerman, 2015; Sen et al., 2008). Our paper extends earlier studies by showing how individual and contribution characteristics may shape the restrictions imposed on contributions, which may in turn have implications for how firms can leverage these contributions (Fosfuri et al., 2008; Nagle, 2018).

Second, our paper contributes to the growing literature on property rights in digital markets (e.g., Boudreau, Jeppesen and Miric, 2022; Nagaraj, 2018; Waldfogel, 2012). Many studies of digital property rights focus on the use of analog rights in a digital world (Lessig, 2008). For instance, patents, copyright and trademarks were intended to protect physical products, but were adapted to digital innovations with varying levels of success. On the other hand, *Creative Commons* licenses were designed to adapt to the unique characteristics of digital products, such as the ease of copying these products (Bauer, Franke and Tuertscher, 2016; Goldfarb and Tucker, 2019). This paper is among the first to study the use of *Creative Commons* licenses in digital markets as an important property right, particularly in relation to preventing others from commercializing a product and potentially competing commercially with the original creator of that product.

Finally, our paper also contributes to the literature on user entrepreneurship. Previous studies have documented how some users may develop innovations, but then also turn those into commercial products (Baldwin, Hienerth and Von Hippel, 2006; de Jong, von Hippel, Gault, Kuusisto and Raasch, 2015; Hienerth, 2006; Shah and Tripsas, 2007). For instance, 43% of significant innovations in windsurfing, skateboarding, and snowboarding were developed and commercialized by end users (Shah, 2003). However, while the transition of online innovations from free contributions to commercial products has been documented, the fact that many contributors may restrict the ability of others to commercialize community contributions, has been overlooked. The results of the present paper document how contributors may freely contribute to a community, but then systematically retain greater commercial rights over their contributions through *Creative Commons* licenses, which can provide a potential opportunity for entrepreneurship later on. This paper is the first to our knowledge to study this process among free contributors to an innovation community.

## 2. LITERATURE

### 2.1. Innovation Communities

Innovation communities have been the source of a number of economically important innovations. Prominent examples include open-source software, developed by a variety of individuals who contribute freely to the development of important software components (Shah, 2006; Lakhani and von Hippel and, 2004), or physical products such as sporting goods (Franke and Shah, 2003) or products for children (Shah and Tripsas, 2007). In addition to the importance of these communities on their own, these communities are important inputs for many firms.

There is evidence that firms may pick valuable community innovations and integrate them into future product development or use them as complementary goods for their original product (Jeppesen and Frederiksen, 2006). In the case of server-software for instance, much of the underlying infrastructure used by commercial companies is based on open-source projects such as *Apache Server*, which is freely available and relied upon by many companies (Franke and von Hippel, 2003). Previous research has shown that innovators in communities are often able to develop new innovations at a greater pace than many traditional producers, such as firms (Baldwin and Von Hippel, 2011). This is partly because firms themselves may struggle to identify and adapt to consumer needs as quickly as individual community members that develop these innovations to suit their own needs. As a result, innovation communities have significant implications not only for the development of novel innovations but also for the welfare of firms and societies (Gambardella et al., 2016).

Previous research has highlighted that two key processes are particularly important for the success of innovation communities. The first key determinant of success is the motivation of individuals to contribute to these communities (Ren et al., 2007; Wiertz and de Ruyter, 2007). Communities require a critical mass of contributors to participate; this encourages others to perceive the benefits of the community and participate as well (Wasko et al., 2009). In addition, communities benefit from having a wide range of different participants and sustaining the engagement of these individuals (Frey and

Gallus, 2017; Dahlander, Jeppesen and Piezunka, 2019). As a result, a key concern in the literature on innovation communities is understanding the factors motivating individuals to contribute to these communities, and the factors that affect their contributions. We review these in detail in *Section 2.2.*

The second key driver of success of communities is the recombinant or cumulative nature of innovation in these communities. A key premise of the open-source movement is that by making code or other products openly available, they could serve as inputs for others' development efforts. Similarly, in many of the innovation communities mentioned above, which developed novel physical products that were later commercialized by companies, a key step in the development of these innovation processes was that these products were freely revealed by innovators and then built upon by others. This cumulative nature of innovation is well established (Murray and O'Mahony, 2007). The increased malleability of information in digital settings creates a vast set of potential reuses of a given innovation and thus provides even more potential benefit from recombination. Because of this malleability, digital products might acquire functions for which they were not originally intended. In the innovation literature, this phenomenon is referred to as exaptation (Andriani and Cattani, 2016). More generally, the recombinant or cumulative nature of innovation is also a critical aspect of how innovation communities function, as contributing on the basis of products created by others has been shown to be a critical driver of individual contributions in a number of studies (Miric, Ozalp and Yilmaz, 2023; Yilmaz, Naumovska and Miric, 2023).

## 2.2. Motivations to Contribute to Innovation Communities

There are two broad categories of motivations behind why individuals contribute to innovation communities: *intrinsic motivations* and *extrinsic motivations*. Intrinsic motivations are those that are related to an "individual's desire to perform the task for its own sake" (Bénabou and Tirole, 2003, p. 490), but do not necessarily translate into direct rewards. For instance, survey-based research on the motives of programmers' participation in open software development projects points to intrinsic motivations resulting from enjoyment of intellectual challenges and their completion (Lakhani and Wolf, 2005).

On the other hand, individuals can also be driven to contribute by extrinsic motivations, i.e., "conditional rewards" (Bénabou and Tirole, 2003, p. 490). For instance, extrinsic motivations include expected benefits from reciprocity, the hope to benefit from one's own contributions, recognition by peers or gains in social capital (Gambardella et al., 2016; Lakhani and Wolf, 2005; Lerner and Tirole, 2002; Wasko and Faraj, 2005)[3]. Specifically, previous research has shown that the prospect of gaining reputation within the community is indeed a key driver for individuals to create (high-quality) contributions (Lakhani and von Hippel, 2004; Lakhani and Wolf, 2005; Wasko and Faraj, 2005)[4].

At the same time, many contributors are able to collect (indirect) monetary rewards, even if they contribute to the community for free. This is the case, for instance, if establishing a reputation as a highly skilled developer in an OOS community leads to job market opportunities outside the community (Lerner and Tirole 2002). Xu, Nian and Cabral (2022) study individual contributors to the *StackOverflow* developer community and find that individuals contribute more to online communities in the period prior to searching for a new job, but decrease their contributions afterwards.

At the same time, in addition to contributing to a community, many individuals may choose to sell contributions from the community as commercial products outside of the community. Many firms that contribute to open-source software development (e.g., Redhat and Oracle), will also offer those products and services to commercial clients (Fosfuri et al., 2008; Nagle, 2018). As a result, being able to use contributions of these communities and convert them into products which may be sold to consumers, is also an important motivator within many communities.

---

[3] Note that the distinction between intrinsic and extrinsic motivations is not always clear-cut. Previous studies have therefore highlighted that many motivations consist of a "varying degree of both intrinsic and extrinsic motivations" (Belenzon and Schankerman, 2015, p. 799)

[4] While some scholars have focused on the motivational effect of potential gains in *reputation* (Lakhani and von Hippel, 2004; Lakhani and Wolf 2005), others have focused on potential gains in the related (but distinct (George, Dahlander, Graffin and Sim, 2016; Washington and Zajac 2005)) concept of *status* (Goes et al., 2016; Roberts et al., 2006) or both (Wasko and Faraj, 2005).

## 2.3. Property Rights in Online Communities

A long-standing puzzle in the literature on innovation communities was why individuals may contribute to such communities, even if they were not able to protect their contributions and others could imitate them easily (Lerner and Tirole, 2001). The key insight from this literature was that individuals would contribute their efforts, but they relied on indirect benefits such as enhanced reputation (Dahlander et al., 2019; Faraj, Kudaravalli, Wasko, 2015; Lappas, Dellarocas and Derakhshani, 2017). However, this requires the existence of property rights which allow individuals to be recognized and collect credit for their contributions. In other settings, these property rights may take the form of legal protections such as patents, copyright or trademarks (Boudreau et al., 2022; Graham, Merges, Samuelson and Sichelman, 2009). However, in a large number of settings, norms and more generative property rights such as *Creative Commons* licenses are used as a way to provide individuals with rights over their creations (Bauer et al., 2016). These property rights provide a way for individuals to freely share their contributions with other members of the community, while retaining control of how they are used, reused or even commercialized. Individuals can allow others to use their contributions, while restricting them, for instance, from reusing (i.e., building upon) these contributions to create other contributions. More commonly, individuals can allow others to use or even reuse the contributions they create, but not allow others to commercialize them.

Previous research has shown that members of innovation communities do indeed care about the property rights related to their innovations, as well as the resulting possibility of using these innovations for commercial purposes (Belenzon and Schankerman, 2015; He, Puranam, Shrestha and von Krogh, 2020; Sen et al., 2008; Singh and Phelps, 2013). Particularly in the context of open-source software projects however, research on licensing choices has often looked at decisions at the project level, i.e., decisions at the level of groups of multiple innovators (e.g., Sen et al., 2008; Singh and Phelps, 2013). Even if a large share of the individual members of innovation communities might not protect their innovations through intellectual property rights, innovation communities a whole, may protect themselves from potential threats by using a number of different tools, including licenses and normative tactics (O'Mahony, 2003). These mechanisms are aimed, for instance, at protecting communities from "hijacking" (Lerner and Tirole, 2005) where firms appropriate innovations from these communities and use them for commercial purposes. Thus, licenses are important from the point of view of firms as well, as they can potentially limit their ability to use innovations (or elements thereof) from the community for commercial purposes (Dahlander and Magnusson, 2005).

In contrast to licensing choices at the level of groups of multiple innovators (e.g., Belenzon and Schankerman, 2015; Sen et al., 2008), licensing choices at the level of individuals within innovation communities have received far less attention. This is a limitation, as group-level licensing choices cannot capture potential effects of restricting commercialization by individual members of the communities and because it may mask heterogeneity among individual members in their propensity to choose certain licenses.

## 3. THEORETICAL FRAMEWORK

We theorize about individuals' choice of non-commercial licenses by focusing on the two key processes within innovation communities that we introduced above, i.e., individuals creating contributions and individuals building on each other's work.

A key mechanism fostering contribution in innovation communities is the ability of individuals to gain reputation within the community (Dahlander et al., 2019; Lappas et al., 2017), as this serves as an important source of (community-driven) extrinsic motivation for these individuals. Higher reputation, defined as "beliefs or perceptions held about the quality of a focal actor" (George et al., 2016, p. 1), can result from the recognition of the quality of the focal individual's contributions, as well as from a greater number of other individuals using, reusing or commercializing the focal individual's creations. At the same time, reputation is also potentially linked to individuals' choices regarding more or less restrictive licenses and the benefits and costs associated with using such licenses.

Allowing others to reuse one's creation or even use it for commercial purposes has been shown, in various settings, to enhance overall demand and awareness. For instance, previous research has shown that reuse can have a positive effect on demand for the original contribution (Watson, 2017) due to advertising effects and increased diffusion, and that this effect is particularly strong for individuals with low reputation. In the context of the music industry, lessening an album's sharing restrictions can increase sales by 10% on average, but this effect is much larger for less popular albums and significantly reduced for top-selling albums (Zhang, 2016). Similarly, Watson (2017) shows that the release of a derivative (or reused) song to the market increases the demand for its upstream product by 3% and that this effect is particularly strong if a song by a less prominent artist is remixed. More generally, if creations are reused, this can facilitate discovery of the original product (Yilmaz et al., 2023), while restricting reuse may limit such discovery effects (Kretschmer and Peukert, 2020; Peukert, Claussen and Kretschmer, 2017). As a result, imposing more restrictive licenses may reduce awareness, which may in turn harm particularly those contributors who do not have an established reputation. Therefore, we might expect individuals with lower reputation to be more likely to use permissive licenses that do not impose constraints on others, such as restricting commercialization.

Moreover, allowing commercial use is more consistent with the implicit norm in many innovation communities that contributions are shared freely and exchanged within the community (Bauer et al., 2016; Franke and Shah, 2003). Research has shown that these expectations shape the behavior of community members who are often driven by the "desire to conform to the norms of the community" rather than an intrinsic interest in ultimate value of their contributions for the community (Shah, 2006). Individuals who do not have an established reputation are likely the ones who benefit the most from conforming to these norms, as this can help them gain reputation in the community. On the other hand, as community members become more established, the potential marginal increase in reputation from conforming to these norms becomes smaller, which reduces the benefit they can derive from choosing less restrictive licenses.

At the same time, the cost of using non-commercial licenses increases with higher reputation. Without imposing restrictive licenses, others could commercialize their contributions or creations that are based on them, possibly even barring original innovators from commercializing their own contributions. This can lead to substantial opportunity costs, particularly for innovators with an established reputation, as they miss out on potentially high financial gains from commercializing their creations. The reputation of individuals is often derived from "delivering quality over time" (George et al., 2016, p. 1), thus suggesting that the potential value of their contributions is higher. Moreover, greater reputation brings more visibility, increasing the chance of their contributions being reused or commercialized in the first place. For instance, Josef Prusa, who is a high reputation innovator in the 3D printing community, received increasing attention for his open-source 3D printers but was also confronted with an increasing number of firms who started selling these printers for a profit. As he did not condone these activities and wanted to secure the position of his own firm, *Prusa Research*, he decided to impose a novel and more restrictive type of license on newer versions of his 3D printers (Stevenson, 2023).

High reputation innovators may also bear psychological costs if they fail to restrict commercialization. Studies have shown that innovators often develop a strong sense of ownership for their creations, especially when they invest significant effort into them (Franke, Schreier and Kaiser, 2010; Moreau, Bonney and Herd, 2011), and perceive themselves as competent in a given task (Furby, 1991; Williams and DeSteno, 2008). High reputation innovators may thus view their creations as a reflection of their skills (Reb and Connolly, 2007) and adopt territorial responses like restrictive licenses if they feel that their creations are threatened (Kirk, Peck and Swain, 2017).

More broadly, as reputation mechanisms in user communities enhance extrinsic motivations, they may impair intrinsic motivations, which is consistent with findings in previous literature (Bénabou and Tirole, 2003; Deci and Ryan, 1985; Jeppesen and Frederiksen, 2006). For instance, the study by Gneezy and Rustichini (2000) shows that volunteers who received a small fraction of the donations they gathered from donors gathered less money overall than those who volunteered for free. Similarly, a study by Jeppesen and Frederiksen (2006) shows that individuals who were more intrinsically motivated were more likely to share their innovations overall. Similar patterns may play out in relation to license choices. As individuals have a more established reputation, they may increasingly shift towards pursuing financial benefits. This in turn may reduce the likelihood of conforming to norms of complete openness, and instead make them restrict commercialization of their contributions to the community.

Taken together, this suggests that as individuals' reputation grows, they will likely utilize more restrictive licenses that prevent others from commercializing their contributions to the community.

**H1.** *Higher reputation contributors are more likely to restrict others from commercializing their contributions.*

As we discussed above, individuals building on each other's work is another key process within innovation communities. Contributions may therefore be either created from scratch or by reusing (i.e., building upon) another contribution. Whether a particular contribution is built from scratch or is based on another contribution may also shape the decision of an individual contributor to use a restrictive license.

In principle, there may be reasons to expect that contributors that reuse existing components are likely to use less restrictive licenses. For instance, communities typically rely on norms around reciprocity, which can create a community-based (extrinsic) motivation to reciprocate by contributing to the community in a similar way as others (Faraj and Johnson, 2011; Ren et al., 2007). If individual contributors impose more restrictive licenses on their products created by remixing and reusing the contributions of others that were available without restrictions, this could be seen as a violation of community norms. Individuals who build on top of existing contributions, may therefore be less likely to impose restrictions on their contributions in an effort to comply with the norm. In a similar vein, individual contributors that reuse existing contributions may feel a sense of indebtedness, in that they need to allow others to build upon and perhaps even commercialize their contributions, as they have built upon the contributions of others.

On the other hand, we argue that there are several factors that will outweigh these mechanisms and will ultimately lead innovators to use more restrictive licenses when building upon the contributions of others. Individual contributors who reuse existing contributions often significantly improve these, transforming incomplete or unrefined components into polished and potentially commercially viable innovations. Research has shown that creations in innovation communities are indeed often incomplete and benefit from being reused and improved by others (Haefliger, von Krogh and Spaeth, 2008; Hill and Monroy-Hernández, 2013). As a result, innovators that take existing contributions and reuse them, may want to impose non-commercial licenses to ensure that the resulting, and potentially more valuable, contribution is protected. In contrast, individuals that create contributions from scratch may face fewer concerns about commercialization due to the unique and potentially unrefined nature of their creations. Consequently, they may be more willing to reduce restrictions, thereby allowing wider reuse and even commercialization of their contributions. Their motivations may be more closely aligned with fostering collaborative innovation than with limiting the commercial use of their work.

In addition, recent studies have shown how products based on the same underlying technology tend to be more similar to each other (Miric et al., 2023). As a result, users may have greater incentives to use more restrictive licenses when building on existing contributions in order to protect their own contributions to the community from competing products that might be built using the same existing contributions. More broadly, when individuals build on the contributions of others, they might be more aware of the fact that others can build on, or even commercialize, their innovations too, which in turn might push them to use restrictive licenses.

Thus, even if innovators might feel some sense of reciprocity towards the community, they will likely want to protect their specific contribution when they choose to reuse an existing contribution.

> **H2.** *Contributors who build on the contributions of others, are more likely to restrict others from commercializing their contributions.*

# 4 DATA AND EMPIRICAL STRATEGY

## 4.1. Empirical Context

We study the use of restrictive (non-commercial) licenses on designs released into the *Thingiverse* 3D-printable design community. The term 3D printing (or additive manufacturing) refers to a manufacturing process where physical products are created by joining materials layer upon layer based on digital 3D-printable designs. Individuals can create and contribute 3D-printable designs to *Thingiverse* and share these designs with others. *Thingiverse* is one of the biggest 3D-printable design communities, owned and operated by *Makerbot,* a large producer of 3D printer devices.

This empirical context is particularly suitable for our study because it tracks individual contributions (i.e., 3D-printable designs), as well as the use of (different types of) *Creative Commons* licenses on these contributions. When individuals share their 3D-printable designs on *Thingiverse*, other users are free to download and 3D-print these designs. However, when individuals release their designs, they also specify whether others can reuse these designs (i.e., build upon the design to create other 3D designs, so-called derivative designs) or commercialize these designs (i.e., commercialize the focal design or derivative designs). If creators restrict commercialization of their designs, this means that, even though the design may be available for free within the community (and can be downloaded and 3D-printed), the creator retains the right to prevent this design or its derivatives from being used for commercial purposes (e.g., selling the 3D design, its derivatives or 3D-printed versions of the two for a profit). The creators however can themselves commercialize the design.

In contrast to other open communities where users contribute to ongoing projects, designs in the *Thingiverse* community are mostly the result of individual efforts, which strengthens the sense of ownership. In the past, there have been cases of infringement of the licenses imposed on *Thingiverse* contributions, which has led to litigation or enforcement of these licenses. An example of this is the case of an *eBay* seller called just3dprint, who has downloaded over two thousand designs from *Thingiverse* and sold them for profit. Many designers in the community expressed complaints about designs being sold for profit, even though the licenses did not allow it. The first notice of this issue came from Louise Driggers, a designer who expressed dissatisfaction with the situation. Subsequently, numerous other designers checked just3dprint's *eBay* page and discovered that their work had also been affected. A significant number of community members sent their complaints to both *eBay* and just3dprint. In response, *Makerbot* also issued a statement regarding the violation of their Terms of Service and mentioned that they would consult their legal team to determine the next steps. Due to the increasing number of complaints filed by both designers and *Makerbot*, *eBay* removed the items in question.[5]

Furthermore, *Makerbot* itself has faced criticism for allegedly commercializing community-created 3D printing tools. The controversy arose when *Makerbot* patents surfaced online, resulting in accusations that the company was appropriating intellectual property from its innovation community by utilizing designs that restricted commercialization (Biggs, 2014). This case highlights the significance for companies to consider the licensing choices of individuals who may be reluctant to relinquish their intellectual property rights.

**4.2 License Types**

*Creative Commons* (CC) licenses are a common way for individuals to release and protect their contributions for online communities. Such licenses are commonly used when contributors release open-source software, images and online photographs, digital designs and self-published books, as well as music and sound effects (Carroll, 2006; Moilanen, Daly, Lobato and Allen, 2014). Approximately 98% of the designs in our empirical setting are licensed under *Creative Commons* licenses. *Creative Commons* licenses are based on combinations of different modules that each specify certain aspects of how the licensed product can or cannot be used. These modules are "Attribution" (which imposes a constraint that follow-on users must acknowledge or cite the original source), "Non-Commercial" (which indicates that follow-on users cannot use both the product and its derivative products for commercial purposes), "Share Alike" (which indicates that the follow-on product must be shared under the same terms as the product being reused), and "No Derivative" (which indicates that this product cannot be used to create derivatives). Individuals can choose between multiple different license types that are based on combinations of these modules, which can effectively reflect the degree of openness of the product. In addition to *Creative Commons* licenses, users can also choose other licenses such as *GNU*, *BSD Licenses* and *Nokia* (see *Figure 1* for the distribution of the licenses used on *Thingiverse*). In *Appendix B*, we provide additional information on the licenses that are used on *Thingiverse*, as well as an overview of how we use these licenses to generate our outcome variables of interest.

In our analysis, we focus on whether the creators impose a *Creative Commons* license that contains the "Non-Commercial" module. These licenses are enforceable, often by the platform that would force the offending party to comply with the licenses, or in some cases have been enforced through litigation.[6]

### 4.3. Data and Sample

Our initial sample consisted of data on 244,990 designs released between 2014 and 2016, published in any of the 79 design sub-categories available on *Thingiverse*. For each design in the sample, we observe the design summary that includes the name and description of the design, the designer of the design, the list and dates of previous designs of the designer, the number of downloads, likes, 3D prints and views, the submission date and the licensing choice of the designer. We also observe the number of followers that each individual has on a weekly basis. We omitted 372 designs with broken printing files, 39,897 derivative designs whose parent designs required derivatives to comply with the licensing terms of the parent designs, and 22,114 designs belonging to users with a single design. Furthermore, we excluded the observations of *MakerBot* (154 observations), which is the owner of the *Thingiverse* platform. Therefore, our final dataset comprises 182,453 designs from 30,093 designers.

---

[6] The *Creative Commons* association maintains a list of example cases where licenses have successfully been enforced through litigation: https://legaldb.creativecommons.org/cases/15/

## 4.4. Outcome Variable: Use of Non-Commercial Licenses

*4.4.1. Use of Non-Commercial License (0/1).* is defined to one if a license for a particular design ($i$) includes the "Non-Commercial" module in the license. This can also be paired with other modules such as "No-Derivative" or "Share Alike" modules. We also validate the analysis in *Sections 5.3.3* and *5.3.4*, where we use alternative outcome variables that take into account other modules.

## 4.5. Independent Variables

*4.5.1. Number of Followers.* We measure the reputation of a particular designer using the *Number of Followers* at the time of a product release. *Thingiverse* allows users to "follow" updates from other users. Generally, users are followed if they are known for making interesting and high-quality contributions, and therefore other members of the community would like to be informed when they have released a new product. A higher number of followers is therefore believed to closely reflect the overall reputation that a user has for providing high-quality contributions[7].

---

[7] The number of followers is a suitable measure for the reputation of an individual as it is the result of "delivering quality over time" (George et al., 2016) and "refers to a summary categorization of real or perceived historical differences in product or service quality" (Washington and Zajac, 2005). It is important to note that reputation is distinct from status and popularity. Status refers to a "socially constructed, intersubjectively agreed-upon and accepted ordering or ranking of individuals" (Washington and Zajac, 2005) that "flows through associations" (George et al., 2016) rather than through previous performance. Popularity typically denotes either "the prevalence or number of prior adoptions of a product" or its "widespread liking" (Kovács and Sharkey, 2014).

*4.5.2. Derivative of Existing Design.* Reuse (i.e., building upon an existing contribution) is a key aspect of innovation in many communities (Haefliger et al., 2008; Hill and Monroy-Hernández, 2013; Yilmaz et al., 2023). Within our setting, approximately 15.5% of all designs are created by modifying existing designs available on the platform. We use the indictor variable *Derivative of Existing Design* to indicate whether a design (i.e., a contribution) is created by reusing existing designs. We also created the variable *Number of Previous Derivatives*, which indicates the number of designs the focal designer has previously created that were based on reusing existing designs. We include this variable as an additional control variable in some of our models to account for potential differences in the total number of derivative designs a designer has created. We log transform the variable. When calculating the variable, we consider the entire stock of designs and derivative designs of users. In other words, we also consider derivative designs that were introduced by the users prior to our starting date in 2014.

*4.5.3. Control Variables.* We include a number of control variables in the analysis. The variable *Market Tenure* captures the number of weeks, log transformed, that an individual designer has been present in the market (i.e., on *Thingiverse*). Furthermore, this variable also enables us to control for designers' tenure-related differences in licensing choices. More experienced designers might also be more informed about the copyright choices available to them and make an effort to set them according to their preferences, while new designers are less likely to pay attention to these details. The variable *File Size,* log transformed in kilobytes, reflects the size of the design file which partly reflects the complexity of the design (Yin, Davis & Muzyrya, 2014), as well as the effort put into the design.

In *Table 1*, we report descriptive statistics and correlations between our variables.

## 4.6. Empirical Strategy

In our analysis, we estimate the correlation of variables capturing contributor reputation and indicators whether her contributions reuse other contributions (i.e., whether they are derivatives), and the decision to use *Non-Commercial* licenses. We estimate the following basic model.

$$\text{Use of Non} - \text{Commercial License}_{jitk}$$

$$= \beta_1 \, \text{Number of Followers}_{jt} + \beta_2 \, \text{Derivative of Existing Design}_i$$

$$+ \text{Controls} + \delta_j + \lambda_k + \gamma_t + u_{it} \qquad (1)$$

We estimate the coefficient at the level of each design ($i$) released by each designer ($j$) in any given period ($t$) as part of design category ($k$). Each design ($i$) released within these categories has unique characteristics influenced by the specific requirements and constraints of the category. The categories include 3D Printing, Art, Fashion, Gadgets, Hobby, Household, Learning, Models, Tools, and Toys & Games. We estimate the results using a correlated random effects probit model. This approach is well suited for panels with binary outcome variables where we attempt to estimate average partial effects (Altonji and Matzkin, 2005; Wooldridge, 2010). This is particularly suitable (shown to be superior to fixed effects and pure random effects) when the number of periods is fewer than eight (Greene, 2004). This is appropriate in our context as we have 5.65 observations per designer on average. Therefore, our main results are correlated random-effects-probit results. However, we also report the results with a linear probability model (LPM) and fixed effects logit model.

In order to attempt to account for unobserved differences between designers ($j$), we follow the approach described by Wooldridge (2010; Chapter 16) and included the averages of each covariate at the designer level. The coefficient of $\delta_j$ indicates the average value of each covariate for each designer, which effectively acts as a fixed effect but is referred to as a correlated designer random effect for each individual designer ($j$). This aspect of the model effectively controls for all time-invariant characteristics at the individual level. These can include traits such as the innate skill level of a designer, their idiosyncratic preferences for certain design features or licenses, or other unobserved characteristics that are consistent over time but may vary between individuals. By controlling for these individual fixed effects, the model mitigates the possibility that these characteristics might confound the relationships we are studying.

We also control for the design category that the design belongs to in order to account for any overall differences in the likelihood of choosing a particular regime at the category level ($\lambda_k$). This helps account for inherent, time-invariant differences between different design categories. Some categories might be more prone to certain licensing regimes due to their inherent characteristics. Controlling for these fixed effects helps ensure that any observed differences in licensing are not simply due to these stable category-specific factors.

We include time period - cohort fixed effects (month and year of design release) indicated by $\gamma_t$. These control for any overall trends or patterns that might affect all designs released at a particular time. For example, if there is a general shift towards more restrictive licenses over time, not controlling for this could potentially bias our results. By including these fixed effects, we account for such systematic, time-specific (but constant within a given period) trends or events.

## 5. RESULTS

### 5.1. Main Results

In *Table 2*, we report the results from the correlated random effects probit model. Columns (1), (3) and (5) contain the coefficients from the probit regressions with our variables of interest, while columns (2), (4) and (6) contain the corresponding marginal effects. From the first row we see that the coefficient for *Number of Followers* is positive and significant (*Column 1*: $\beta$ = 0.0356; *S.E.* = 0.0144; $p$ = 0.0130). This is consistent across specifications. The coefficient on the *Number of Followers* suggests that a 100% increase in the number of followers (effectively doubling the number of followers) corresponds to a 10 percentage point increase in the likelihood of using a non-commercial license (column (2)). In columns (3) and (4), we introduce *Derivative of Existing Design (0/1)* as an additional variable. We find that the relationship is positive and significant (*Column 3*: $\beta$ = 0.0980; *S.E.* = 0.0162; $p$ = 0.0000). The results indicate that if a design builds on another design, this increases the likelihood of using a non-commercial license by 2.94 percentage points. It is important to note that the baseline rate is 0.15, and therefore this corresponds to an increase of 19.6% (0.0294/0.15). These results collectively offer preliminary support for Hypotheses 1 and 2. Interestingly, we find no evidence for a statistically significant association between *Number of Previous Derivates* and our outcome variable. This suggests that, while building on other contributions shapes individuals' choice of license, this is mainly true if the focal design is a derivative design, while the number of previously created derivatives seems less relevant.

## 5.2. Control Function Approach

A potential concern with our baseline specification is the potentially endogenous relationship between reputation and the choice of license type. This may be influenced by time variant unobserved factors which can be correlated with both reputation and license type. To account for the unobserved factors that might affect the decision to choose a particular license, we use a control function (CF) approach.

We exploit design promotions performed by *Thingiverse* as an instrument that directly affects the reputation of developers but is not directly linked to their decision to use a particular license. These promotions take the form of featuring individual designs prominently (i.e., with a larger picture and description) on the main page of the *Thingiverse* website and labelling them as being "featured". There is rich evidence on how platforms may selectively promote certain complements, which creates benefits for specific complementors such as increasing their visibility and popularity (Foerderer, Lueker and Heinzl, 2021; Rietveld, Seamans and Meggiorin, 2021; Rietveld, Schilling and Bellavitis, 2019). In our setting, while designs may be featured, these are very rare events and unexpected from the side of the individual designers. There is little evidence that this may be strategically acted upon by designers because it is such a rare event, and many are surprised that they are ever selected. For instance, one designer posted the following after being featured: "*Featured! This is crazy! I told my daughter and she's as blown away as I am.*" Another designer said: "*... It is just so amazing being featured on Thingiverse. When I woke up and saw the e-mail with the notification I couldn't believe it*". Being featured also substantially increases designers' reputation and visibility within the community. In *Figure 2*, we report the average number of weekly new followers that individuals receive before and after their designs are featured. As the results indicate, there is a substantial and sudden increase in the number of new followers after being featured.

The CF approach we take is analogous to the instrumental variable approach implemented with two-stage least squares (2SLS) (Ebbes, Papies and Van Heerde, 2011; Guo and Small, 2016; Petrin and Train; 2010, Wooldridge, 2015; York, Vedula and Lenox, 2018). When the first and second stages are linear, the CF approach is identical to the 2SLS. However, the CF approach is better suited when the second stage outcome variable is discrete (Guo and Small, 2016; Wooldridge, 2015; York et al., 2018). In the CF approach, we estimate a first stage similar to the first stage of an 2SLS model and regress the endogenous variable on covariates and our instrumental variable. Next, we predict the residuals from this first stage regression. Then in the second stage, we include these residuals as an additional regressor along with the endogenous variable and run the probit model. By introducing this covariate, we account for the potential unobserved effects of selection, and when it is included in the model, the remaining coefficients reflect unbiased estimates.[8] Additionally, we bootstrap the standard errors 50 times.

In *Table 3*, we report the results using the control function approach, where we exploit the fact that some designs may be featured by the platform. In columns (1), (2) and (3), we report the results of the first stage regression. We find that being featured is associated with a 183% increase in the number of followers (*Column 3*: $\beta$ = 1.8290; *S.E.* = 0.1723; $p$ = 0.0000). This is because the average designer has on average a small number of followers, and being featured provides a considerable rise to followers. In columns (3), (4), and (5), we report the results for the corresponding instrumented regressions. The coefficient for *Number of Followers* remains positive and significant (*Column 6*: $\beta$ = 0.1607; *S.E.* = 0.0383; *p* = 0.0000).

---

[8] The most well-known example of Control Function methods are Heckman Selection Correction methods. The approach introduced is equivalent to introducing the inverse-mills ratio into the regression and evaluating the remaining coefficients.

While it is reasonable to assume that designers do not anticipate being featured, one remaining concern is that featured designers may inherently differ from the designers in our sample whose designs are not featured. Although we control for time-invariant characteristics of designers by including designer fixed effects (or covariate averages), there is a potential endogeneity issue if designers who choose more open licenses are more likely to be chosen. This could undermine the validity of "*Featured*" as an instrument.

To address this concern, we conducted additional analyses using Coarsened Exact Matching (CEM) to match award-winning designers with those who do not win an award based on pre-award characteristics. These characteristics include the number of previous designs with licenses that contain the "Non-Commercial" module, the number of previous designs with licenses that contain the "Non-Derivative" module, the number of previous derivatives, and the number of followers. We instructed the command to identify matches using five equally spaced bins for each variable. We ensured an exact match on the cohort, meaning that these designers should have similar characteristics in the month when the award was given. We implemented one-to-one matching, selecting a control designer for each designer who won an award. The results from these analyses are reported in *Table A1* in the *Appendix* and are largely consistent with the baseline analyses.

### 5.3. Alternative Specifications and Robustness Checks

We conduct a series of robustness checks and supplementary analyses to validate and expand our findings.

*5.3.1 Linear Probability Regressions*

We report the results using a linear probability model (OLS with a binary outcome variable) and re-estimating the regression results in *Table 4*. Column (1) of *Table 4* includes results from a simple OLS regression. Throughout columns (2)-(8), we sequentially add designer, cohort and category fixed effects. As the adjusted R-squared statistics across different columns indicate, most of the variation in the licensing choice is explained by user fixed effects, followed by category fixed effects. Across different models, the coefficient estimates are comparable with the ones from the probit model.

*5.3.2 Fixed Effects Logit Regressions*

In the main results, we utilized a random effects model, while including individual specific averages in order to capture unobserved individual level effects (Altonji and Matzkin, 2005; Wooldridge, 2010). As an alternative specification, we repeat the analysis by using a logit model with individual-level fixed effects. We report the results in *Table 5*. Given that the fixed effect logit regression drops users with no variation in the dependent variable, the sample size is smaller compared to the main analyses. The overall pattern of results remains unchanged with strong positive association between our outcome variable and both *Number of Followers* and *Derivative of Existing Design (0/1)*.

*5.3.3 Alternative Outcome Variable: Non-Derivative Licenses*

We also explored the consistency of our results to alternative types of restrictive licenses. More specifically, we study the use of licenses that include the *"Non-Derivative"* module, which restricts reuse of the focal design by other designers within the community even if they do not intend to commercialize the original design or its derivative. We replicate all of our analyses with this alternative outcome variable and report the results in *Tables A2* to *A5* in the *Appendix*.

Similar to the results with our main outcome variable, we find evidence for a positive association between *Number of Followers* and the use of *Non-Derivative* licenses, although the effects are not statistically significant in all specifications. Furthermore, our results also provide evidence for a negative association between *Derivative of Existing Design (0/1)* and our alternative outcome variable, which is different from the positive sign we obtained with our main outcome variable.

The difference in what these licenses aim to protect might explain the different results in the analyses. For instance, having a high number of followers might correlate positively with both license types as these high reputation designers face the same potential costs and benefits in both cases, and want to protect their work in either case. However, creating a derivative design could influence these licenses differently. Designers that build upon existing contributions might understand the value of sharing and reusing (within the community), and might therefore be less likely to use a *Non-Derivative* license. On the other hand, designers that build upon existing contributions might still want to prevent commercial exploitation of their own and the original designer's work, which could also be seen as taking something away from the community. Hence, they might prefer to use a *Non-Commercial* license.

### 5.3.4 Alternative Outcome Variable: Closedness of Licenses

The different licenses in our empirical setting can be thought of as representing different levels of closedness (as opposed to openness) regarding the use of contributions shared within the community, with some licenses being more closed (or restrictive) than others. As an alternative outcome variable, we therefore created the ordinal variable *Closedness*, which is equal to 0 if a "GNU Lesser General", "Public Domain" or "BSD License" license has been chosen, 1 if an "Attribution", or "GNU GPL" license has been chosen, 2 if an "Attribution-Share Alike" or "Attribution-Non Commercial" license has been chosen, 3 if an "Attribution-No Derivative", "Attribution-Non Commercial-Share Alike" or "Nokia (3D Printing)" license has been chosen and 4 if an "Attribution-Non Commercial-No Derivative" or "All Rights Reserved" license has been chosen. We use an ordered probit model and an OLS model to examine if our results remained consistent with a broader definition of closedness. The results, reported in *Table A6* in the *Appendix*, demonstrate similarities to the outcomes obtained in our previous analyses. On the one hand, the results suggest that *Number of Followers* is positively associated with licenses that are more closed. This can suggest that, for instance, higher-reputation users tend to also restrict reuse within the community rather than merely restricting commercialization. On the other hand, *Derivative of Existing Design (0/1)* is negatively associated with licenses that are more closed. This can suggest that, for instance, if the focal design is a derivative, designers will tend to merely restrict commercialization but not reuse within the community. The latter result is consistent with the results we obtained when *"No Derivative"* was used as the outcome variable.

The instrumental variable approach, commonly employed to infer causative effects from observational data, can sometimes introduce unnecessary ambiguity into the results despite reducing bias. As such, recent econometrics literature has emphasized the need for researchers to determine whether the bias is substantial enough to warrant the use of the instrumental variable approach. Scholars also recommend conducting sensitivity assessments to gauge the robustness of the results against potential unseen confounding factors (Lal, Lockhart, Xu and Zu, 2021). Such analyses require identifying possible unobserved variables that could drastically alter the interpretation of the estimated causal effect and evaluating the probability of such confounding. The latter is ascertained through the research model and an expert understanding of the data generation process.

In order to quantify the magnitude of the impact of unobserved variables required to nullify the findings, we adopt the methodology proposed by Cinelli and Hazlett (2020). This approach provides a comparative scale grounded on a reference variable, selected through a comprehensive understanding of the context. In our analyses, we used *LnFileSize* as the reference variable, studying how much greater the influence of unobserved variables would need to be compared to *LnFileSize*, a significant predictor of various licensing choices. The findings suggest that the effects would have to exceed the *LnFileSize* effect by more than fifty times to start discrediting the results. The specifics of this technique and our findings are detailed in *Appendix C.*

## 6. DISCUSSION AND CONCLUSION

In this paper, we investigate factors that are associated with the decision of individuals to restrict commercial use of their contributions to innovation communities. We do so by focusing on the two key processes within innovation communities, i.e., individuals creating contributions and individuals building on each other's work, as well as the related incentives for contributors to engage in these processes. We explore how individual reputation is associated with the impact of using non-commercial licenses and find that individuals with greater reputation are more likely to use such licenses. In addition, we find evidence that creating contributions by reusing (i.e., building upon) existing components, is also associated with a higher likelihood of using restrictive non-commercial licenses.

*Creative Commons* licenses are an important way of introducing and enforcing property rights in digital markets. They can serve as a complement to other forms of property rights that may not be as effective in digital markets (Boudreau et al., 2022). While using formal IP may be a more common strategy among firms and professional inventors, *Creative Commons* licenses are commonly used in digital markets to protect products which are being revealed and may be easily imitated or copied. However, the factors associated with the use of these types of licenses in innovation communities were not previously studied.

Our results show how individuals' choice of open or restrictive licenses, such as non-commercial licenses, is associated with reputation. As individuals gain reputation, they become more likely to use non-commercial licenses and constrict the ability of others in the community to commercialize these products. This is consistent with our theoretical arguments that when individuals do not have an established reputation, they may benefit from using more open licenses for their contributions as this can lead to greater visibility and advertising effects. These benefits however may decrease if individuals gain in reputation, while the associated costs from foregoing the opportunity to commercialize their own contributions and from a feeling of possession of their contributions may increase. Therefore, high-reputation innovators may be more inclined to impose non-commercial licenses as these can potentially allow them to capture more value from the contributions they create.

In addition, our results provide some evidence that individuals are also more likely to impose non-commercial licenses on their products when they reuse (i.e., build upon) existing contributions to create new contributions. This is consistent with our arguments that contributions which are created by reusing existing contributions are often more polished, and thus potentially more valuable for commercialization, and more vulnerable to competitors potentially using the same contributions as building blocks to build similar derivative contributions. Therefore, even if individuals may in principle feel that they should reciprocate to others allowing them to reuse their contributions by choosing less restrictive licenses, they might ultimately want to protect their (derivative) contributions through non-commercial licenses. More generally, these results suggest that reusing contributions created by others may raise the awareness of individuals that their contributions might be reused (and potentially commercialized) by others, too.

These decisions about license choice imply that while individuals may freely release their contributions, and allow others to use them, there are important individual decisions regarding the subsequent commercialization of these contributions. These may have important welfare implications for the community, but also for the organizations that expect to benefit from commercializing innovations created in these communities.

## 6.1. Theoretical Implications

These findings provide several important contributions to the literature. First, this paper contributes to the literature on online innovation communities. Much of the prior literature has highlighted how and why individuals may choose to contribute to these communities (Gambardella et al., 2016; Lakhani and Wolf, 2005; Roberts et al., 2006). Yet, the majority of these studies take a static view and consider the motives for individuals to initially begin contributing to these communities. However, as existing studies have shown, these motivations may change with time or as a consequence of events surrounding the community (Miric and Jeppesen, 2023; Nagaraj and Piezunka, 2020). The present paper builds on these studies to show how the way that individuals contribute to the community (or the property rights they impose around their contributions) changes as innovators gain reputation over time and depending on the nature of how the contribution was created. In particular, while previous research has highlighted potential benefits of using reputation-based mechanisms to incentivize individuals to *contribute* to the community (Grant and Betts, 2013; Lappas et al., 2017), our results suggest that this may come at the cost of reducing the probability that individuals allow commercialization of these contributions. This has important implications because the contributions of a handful of "high reputation" contributors may be particularly important to these communities. At the same time, while the decision to impose non-commercial licenses may ensure that the contributions of one individual are not taken outside the community by others and commercialized, it may also lead to less diffusion and use of these contributions overall, which might have further implications for the community itself.

Second, a growing body of literature has considered the use of property rights in digital markets (e.g., Boudreau et al., 2022; Nagaraj, 2018; Waldfogel, 2012). One aspect that has been less studied is the use of *Creative Commons* licenses, which may be particularly suitable for digital products such as software, music or digital designs which can be easily replicated and manipulated. These property rights serve as a way for individuals to retain some ownership and control over their creations, while simultaneously releasing them and making them freely available to the public. The results of our analysis contribute to our understanding of this phenomenon by providing evidence of the factors associated with using this type of property right. In particular, individual reputation and building on existing contributions is associated with individuals being more likely to restrict the ability of others to commercialize their contributions through non-commercial licenses. These findings contribute to the literature on digital property rights by showing how different factors are associated with the use of this type of digital property right.

Lastly, our findings also contribute to the literature on user entrepreneurship. While previous research has shown how some users may turn their innovations into commercial products (Baldwin et al., 2006; de Jong et al., 2015; Hienerth, 2006; Shah and Tripsas, 2007), much less focus has been put on how the transition into commercial products might be shaped by individuals' choices of restrictive licenses for their contributions. These choices might be related both to the intention of preventing others from commercializing these contributions and to the intention of securing the opportunity to commercialize these contributions themselves at a later point in time.

## 6.2. Managerial Implications

Innovation communities are an important source of many societal and business relevant innovations (Franke and Shah, 2003; Shah, 2006; Shah and Tripsas, 2007; von Hippel, 1976). These communities often rely on individual contributors freely sharing their products or contributions, and allowing others to use and build upon these creations. Our results point to the factors that are associated with individuals imposing non-commercial restrictions on the products they create, preventing them from being commercialized by others or being used to create new products which may be then commercialized. This has two important but competing implications. First, from the perspective of contributors and producers, this implies that as individuals gain in reputation, they have more of an incentive to protect their creations and hopefully appropriate value from their efforts. However, on the other hand, many important innovations are taken from these communities and converted into commercial products. The literature on user innovations has a number of examples of important innovations from such communities becoming commercialized and becoming important products (Franke and von Hippel, 2003; Jeppesen and Frederiksen, 2006). These results suggest that particularly high reputation individuals may be unwilling to allow others to commercialize their creations which may prevent some products from moving outside of the community. Understanding the nature of when and which type of *Creative Commons* licenses are used is important for our understanding of how these communities and property rights in digital markets interact.

## 6.3. Conclusions

In this paper, we explore how the reputation of individual contributors in an online community and building on the components created by others, is associated with the types of license individuals choose to protect the products they freely share with the community. Our results indicate how these factors shape the use of non-commercial licenses which we argue has important implications for both our understanding of innovation communities and property rights in digital markets, as well as implications for our understanding of how to manage digital innovation and innovation in online communities.

## 6.4. Limitations

While the analyses presented above provide evidence consistent with our claims, this study also has some limitations. First, the findings from this research may not be applicable to all innovation communities due to the diversity and complexity of different online platforms, as well as the nature of contributions. There might be nuances and factors influencing the decision to use non-commercial licenses that are not captured in this study. Second, while our study acknowledges that both rational factors (e.g., calculating costs and benefits) and emotional factors (e.g., emotional attachment to the contribution) may come into play in decision-making, it does not directly test which mechanism is more effective. Future studies could use surveys to better understand whether rational or emotional factors are more effective in decision-making. Third, we do not consider the role of legal knowledge or awareness in choosing licenses. To account for this, we used individual fixed effects and tenure in the market as controls. However, not all contributors may have a thorough understanding of the licensing options available and their implications.

**TABLES AND FIGURES**

**Figure 1.** Distribution of different licenses on *Thingiverse*

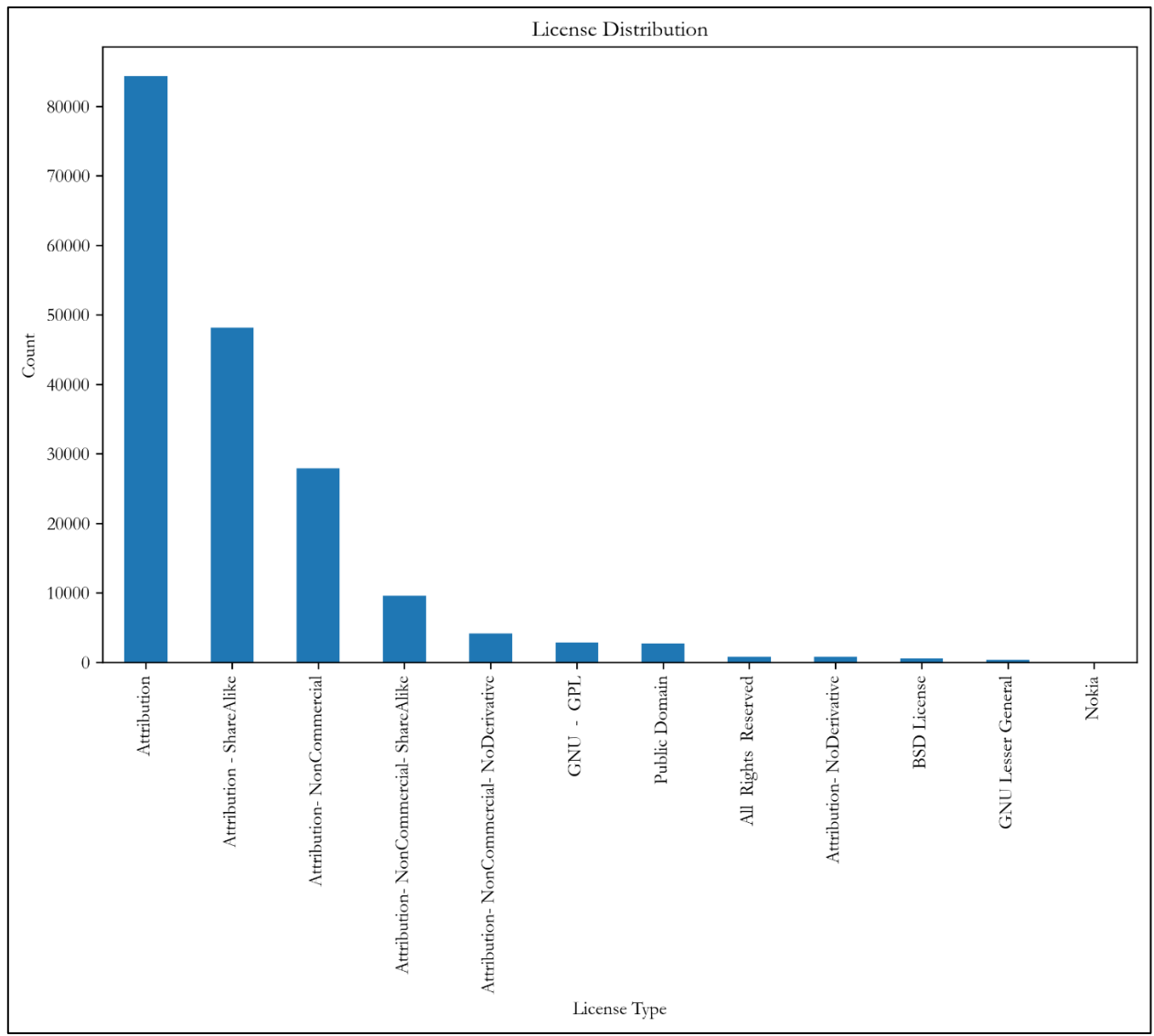

**Figure 2.** Number of new followers before and after being featured on the *Thingiverse* website

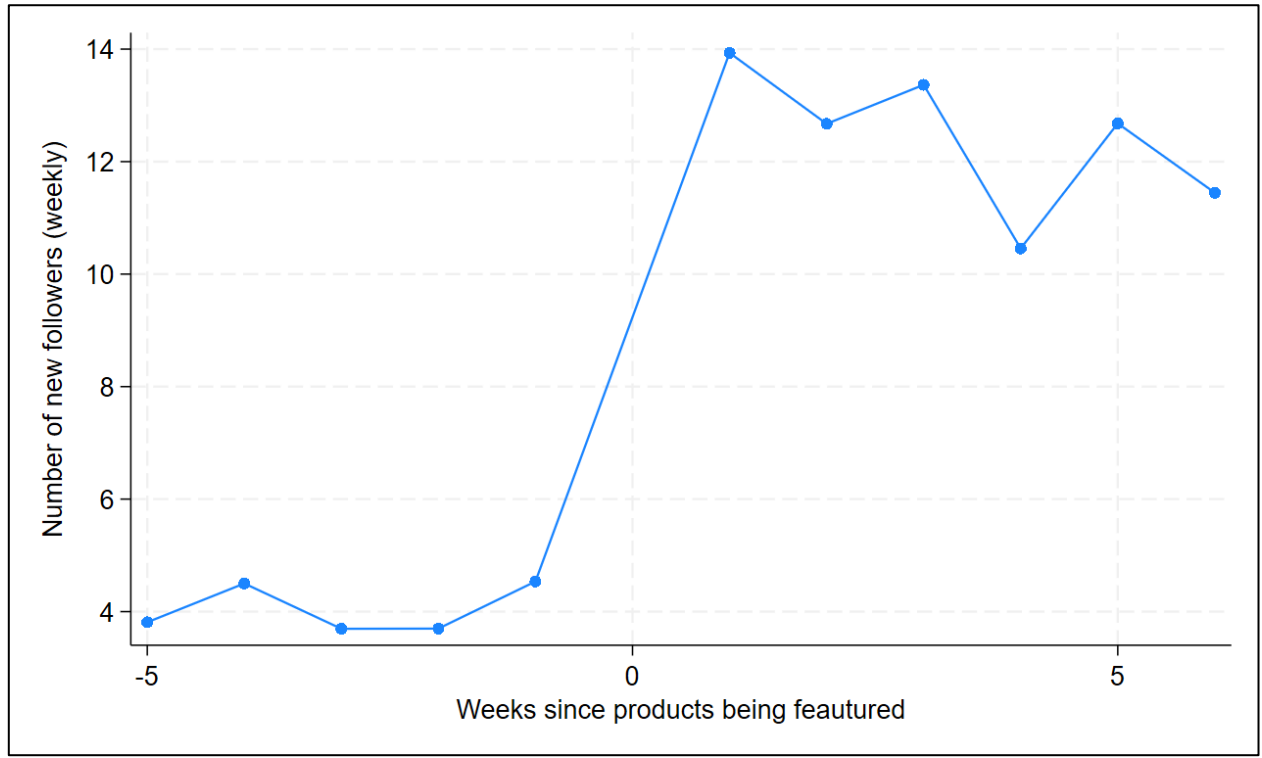

**Table 1.** Descriptive statistics and pairwise correlations

| Variables | Mean | S.D. | (1) | (2) | (3) | (4) | (5) | (6) | (7) |
|---|---|---|---|---|---|---|---|---|---|
| *(1) Non-Derivative License* | 0.032 | 0.175 | 1.000 | | | | | | |
| *(2) Non-Commercial License* | 0.233 | 0.423 | 0.267 | 1.000 | | | | | |
| *(3) Number of Followers* | 1.134 | 1.417 | 0.135 | 0.158 | 1.000 | | | | |
| *(4) File Size* | 12.974 | 2.646 | 0.027 | 0.051 | 0.126 | 1.000 | | | |
| *(5) Number of Previous Derivates* | 0.606 | 1.054 | -0.029 | 0.041 | 0.363 | 0.074 | 1.000 | | |
| *(6) Derivative of Existing Design (0/1)* | 0.155 | 0.362 | -0.066 | 0.025 | -0.051 | 0.056 | 0.403 | 1.000 | |
| *(7) Market Tenure (months)* | 4.935 | 1.804 | 0.042 | 0.066 | 0.399 | 0.106 | 0.313 | 0.086 | 1.000 |

*Note. Pairwise correlation with a significance level exists for all pairwise comparisons in the analysis.*

**Table 2**. Probit Regression Results for Decision to Use Non-Commercial Licenses

**Outcome Variable:** *Use of Non-Commercial License (0/1)*
**Unit of Observation:** *Individual design (i) released by focal designer (j)*
**Model:** *Correlated Random Effects Probit Model*

| | (1) Probit Model Results | (2) Marginal Effects | (3) Probit Model Results | (4) Marginal Effects | (5) Probit Model Results | (6) Marginal Effects |
|---|---|---|---|---|---|---|
| *Number of Followers* | 0.0356 | 0.0107 | 0.0341 | 0.0102 | 0.0399 | 0.0120 |
| | (0.0144) | (0.0043) | (0.0143) | (0.0043) | (0.0143) | (0.0043) |
| | [0.0130] | [0.0133] | [0.0172] | [0.0175] | [0.0053] | [0.0054] |
| *Derivative of Existing Design (0/1)* | | | 0.0980 | 0.0294 | 0.1012 | 0.0303 |
| | | | (0.0162) | (0.0049) | (0.0166) | (0.0050) |
| | | | [0.0000] | [0.0000] | [0.0000] | [0.0000] |
| CONTROLS | | | | | | |
| *Number of Previous Derivates* | -0.0082 | -0.0025 | | | -0.0162 | -0.0048 |
| | (0.0151) | (0.0045) | | | (0.0148) | (0.0044) |
| | [0.5869] | [0.5868] | | | [0.2759] | [0.2755] |
| *File Size* | 0.0220 | 0.0066 | 0.0213 | 0.0064 | 0.0214 | 0.0064 |
| | (0.0022) | (0.0006) | (0.0022) | (0.0006) | (0.0022) | (0.0006) |
| | [0.0000] | [0.0000] | [0.0000] | [0.0000] | [0.0000] | [0.0000] |
| *Market Tenure* | -0.0155 | -0.0046 | -0.0161 | -0.0048 | -0.0140 | -0.0042 |
| | (0.0064) | (0.0019) | (0.0065) | (0.0019) | (0.0065) | (0.0019) |
| | [0.0163] | [0.0160] | [0.0129] | [0.0126] | [0.0307] | [0.0303] |
| *Constant* | -1.0302 | | -1.0194 | | -1.0499 | |
| | (0.1356) | | (0.1358) | | (0.1371) | |
| | [0.0000] | | [0.0000] | | [0.0000] | |
| *Designer Correlated RE* | Yes | Yes | Yes | Yes | Yes | Yes |
| *Cohort FE* | Yes | Yes | Yes | Yes | Yes | Yes |
| *Category FE* | Yes | Yes | Yes | Yes | Yes | Yes |
| *Observations* | 182453 | 182453 | 182453 | 182453 | 182453 | 182453 |
| *Pseudo R²* | 0.0354 | 0.0354 | 0.0365 | 0.0365 | 0.0382 | 0.0382 |

*Note.* Standard errors, reported in the parentheses, are robust to heteroscedasticity and clustered at the user level. p-values are reported in brackets. Probit regression results are reported in Columns 1, 3 & 5. Corresponding marginal effects (at means) are reported in columns 2, 4 and 6. Because the probit coefficients effects are challenging to interpret, the corresponding marginal effects provide a more intuitive interpretation. In Column 2, a one percent increase in the number of followers (log scaled variable) corresponds to a 0.1% increase in the likelihood of using a non-commercial license. Magnitudes are comparable in Columns 3-6. As a result, a 100% increase in the number of followers (effectively a doubling of followers) would correspond to a 10% increase in the likelihood of using a non-commercial license. Results are significant at the 95% level across specifications (Col 1: $\beta = 0.0356$, SE = 0.0144; p = 0.0130; Col 3: $\beta = 0.0341$, SE = 0.0143, p = 0.0172; Col 5: $\beta = 0.0399$, SE = 0.0143, p = 0.0053). Results consistent with H1. In Models $3 - 6$, the variable for whether a particular design is a derivative design is introduced. The results are positive (Col 3: $\beta = 0.0980$, SE = 0.0162; p = 0.0000; Col 5: $\beta = 0.1012$, SE = 0.0166, p = 0.0000) and significant across specifications, consistent with H2. Results indicate that when a design is a derivative design, this would increase the likelihood in using a noncommercial license by 2.94 percentage points (Column 4). It is important to note that the baseline rate of being a derivative is 0.15, and therefore this corresponds to an increase of 19.6% (0.0294/0.15).

**Table 3.** Results for Decision to Use Non-Commercial Licenses, first and second-stage IV regression

**First Stage Outcome Variable:** *Number of Followers.*
**Second Stage Outcome Variable:** *Use of Non-Commercial License (0/1).*
**Instrument:** *Indicator for Period After Contributor is Features in the Community.*
**Unit of Observation:** *Individual design (i) released by focal designer (j).* **Model:** *Control Function Approach*

| Outcome Variable | (1) | (2) | (3) | (4) | (5) | (6) |
|---|---|---|---|---|---|---|
| | First Stage | | | Second Stage | | |
| | *Number of Followers* | | | *Non-Commercial Licensee* | | |
| **INSTRUMENTED VARIABLE** | | | | | | |
| *Number of Followers* | | | | 0.1496 | 0.1530 | 0.1607 |
| *(Instrumented)* | | | | (0.0385) | (0.0367) | (0.0383) |
| | | | | [0.0001] | [0.0000] | [0.0000] |
| *Featured (Instrument)* | 1.9462 | 1.8729 | 1.8290 | | | |
| | (0.1645) | (0.1717) | (0.1723) | | | |
| | [0.0000] | [0.0000] | [0.0000] | | | |
| **REMAINING VARIABLES** | | | | | | |
| *Derivative of Existing* | 0.0148 | | -0.0277 | 0.0962 | | 0.1047 |
| *Design (0/1)* | (0.0082) | | (0.0079) | (0.0119) | | (0.0131) |
| | [0.0725] | | [0.0005] | [0.0000] | | [0.0000] |
| *Number of Previous Derivates* | | 0.3530 | 0.3423 | | -0.0512 | -0.0600 |
| | | (0.0169) | (0.0165) | | (0.0195) | (0.0167) |
| | | [0.0000] | [0.0000] | | [0.0087] | [0.0003] |
| *File Size* | -0.0038 | -0.0042 | -0.0043 | 0.0219 | 0.0227 | 0.0221 |
| | (0.0018) | (0.0018) | (0.0016) | (0.0019) | (0.0019) | (0.0022) |
| | [0.0390] | [0.0173] | [0.0087] | [0.0000] | [0.0000] | [0.0000] |
| *Market Tenure* | 0.2534 | 0.2091 | 0.1973 | -0.0462 | -0.0407 | -0.0383 |
| | (0.0085) | (0.0081) | (0.0078) | (0.0106) | (0.0094) | (0.0086) |
| | [0.0000] | [0.0000] | [0.0000] | [0.0000] | [0.0000] | [0.0000] |
| *Designer Correlated RE* | Yes | Yes | Yes | Yes | Yes | Yes |
| *Cohort FE* | Yes | Yes | Yes | Yes | Yes | Yes |
| *Category FE* | Yes | Yes | Yes | Yes | Yes | Yes |
| *Observations* | 182453 | 182453 | 182453 | 182453 | 182453 | 182453 |
| *Adjusted R²* | 0.353 | 0.372 | 0.432 | | | |
| *Pseudo R²* | | | | 0.036 | 0.037 | 0.039 |

*Note.* Standard errors, reported in the parentheses, are robust to heteroscedasticity and clustered at the user level. p-values are reported in brackets. <u>Instrument and first stage model validity</u>: Given that the reduced form for the endogenous explanatory variable is linear, we use the same diagnostics as in the linear case. The Cragg–Donald Wald F-test statistic rejects the null hypothesis of a weak instrument. The test statistic is 7957.347, which exceeds the critical value of 16.38 proposed by Stock and Yogo (2005). The Kleibergen–Paap rk Wald F-statistics is 7626.608, which also allays concerns over a weak instrument.

**Table 4.** Results for Decision to Use Non-Commercial Licenses, LPM

**Outcome Variable:** *Use of Non-Commercial License (0/1)*
**Unit of Observation:** *Individual design (i) released by focal designer (j)*
**Model:** *Fixed Effects OLS (Linear Probability Model)*

| | (1) | (2) | (3) | (4) | (5) | (6) | (7) | (8) |
|---|---|---|---|---|---|---|---|---|
| *Number of Followers* | 0.0469 | 0.0244 | 0.0470 | 0.0460 | 0.0196 | 0.0198 | 0.0197 | 0.0200 |
| | (0.0008) | (0.0043) | (0.0055) | (0.0051) | (0.0047) | (0.0047) | (0.0047) | (0.0047) |
| | [0.0000] | [0.0000] | [0.0000] | [0.0000] | [0.0000] | [0.0000] | [0.0000] | [0.0000] |
| *Derivative of Existing Design (0/1)* | 0.0369 | 0.0282 | 0.0366 | 0.0421 | 0.0279 | 0.0277 | | 0.0278 |
| | (0.0027) | (0.0050) | (0.0072) | (0.0067) | (0.0050) | (0.0049) | | (0.0050) |
| | [0.0000] | [0.0000] | [0.0000] | [0.0000] | [0.0000] | [0.0000] | | [0.0000] |
| CONTROLS | | | | | | | | |
| *Number of Previous Derivates* | | | | | | | 0.0010 | -0.0014 |
| | | | | | | | (0.0044) | (0.0044) |
| | | | | | | | [0.8274] | [0.7579] |
| *File Size* | 0.0048 | 0.0072 | 0.0048 | 0.0043 | 0.0071 | 0.0067 | 0.0070 | 0.0067 |
| | (0.0004) | (0.0006) | (0.0021) | (0.0020) | (0.0005) | (0.0005) | (0.0005) | (0.0005) |
| | [0.0000] | [0.0000] | [0.0212] | [0.0316] | [0.0000] | [0.0000] | [0.0000] | [0.0000] |
| *Market Tenure* | -0.0006 | 0.0013 | -0.0015 | -0.0020 | -0.0008 | -0.0013 | -0.0014 | -0.0013 |
| | (0.0006) | (0.0016) | (0.0019) | (0.0018) | (0.0018) | (0.0018) | (0.0018) | (0.0018) |
| | [0.3116] | [0.4107] | [0.4203] | [0.2593] | [0.6374] | [0.4612] | [0.4368] | [0.4790] |
| *Constant* | 0.1153 | 0.1015 | 0.1197 | 0.1289 | 0.1180 | 0.1253 | 0.1260 | 0.1256 |
| | (0.0054) | (0.0098) | (0.0269) | (0.0256) | (0.0115) | (0.0115) | (0.0116) | (0.0115) |
| | [0.0000] | [0.0000] | [0.0000] | [0.0000] | [0.0000] | [0.0000] | [0.0000] | [0.0000] |
| *Designer FE* | No | Yes | No | No | Yes | Yes | Yes | Yes |
| *Cohort FE* | No | No | Yes | No | Yes | Yes | Yes | Yes |
| *Category FE* | No | No | No | Yes | No | Yes | Yes | Yes |
| *Observations* | 182453 | 182453 | 182453 | 182453 | 182453 | 182453 | 182453 | 182453 |
| *Adjusted R²* | 0.0270 | 0.580 | 0.0287 | 0.0375 | 0.580 | 0.582 | 0.582 | 0.582 |

Note. Standard errors, reported in the parentheses, are robust to heteroscedasticity and clustered at the user level. p-values are reported in brackets.

**Table 5.** Results for Decision to Use Non-Commercial Licenses, Fixed Effect Logit Model

**Outcome Variable:** *Use of Non-Commercial License (0/1)*
**Unit of Observation:** *Individual design (i) released by focal designer (j)*
**Model:** *Fixed Effects Logit Model*

|  | (1) | (2) | (3) |
|---|---|---|---|
| *Number of Followers* | 0.2551 | 0.2632 | 0.2660 |
|  | (0.0241) | (0.0245) | (0.0245) |
|  | [0.0000] | [0.0000] | [0.0000] |
| *Derivative of Existing Design (0/1)* | 0.2432 |  | 0.2493 |
|  | (0.0286) |  | (0.0287) |
|  | [0.0000] |  | [0.0000] |
| CONTROLS |  |  |  |
| *Number of Previous Derivates* |  | -0.0387 | -0.0593 |
|  |  | (0.0255) | (0.0256) |
|  |  | [0.1287] | [0.0206] |
| *File Size* | 0.0873 | 0.0900 | 0.0875 |
|  | (0.0053) | (0.0053) | (0.0053) |
|  | [0.0000] | [0.0000] | [0.0000] |
| *Market Tenure* | -0.0062 | -0.0047 | -0.0030 |
|  | (0.0130) | (0.0130) | (0.0131) |
|  | [0.6309] | [0.7167] | [0.8205] |
| *Designer FE* | Yes | Yes | Yes |
| *Cohort FE* | Yes | Yes | Yes |
| *Category FE* | Yes | Yes | Yes |
| *Observations* | 72487 | 72487 | 72487 |
| *Pseudo* $R^2$ | 0.0272 | 0.0258 | 0.0273 |

Note. Standard errors, reported in the parentheses, are robust to heteroscedasticity and clustered at the user level. p-values are reported in brackets.

**Table 6**. Probit Regression Results for Decision to Use No-Derivative Licenses

**Outcome Variable:** *Use of No-Derivative License (0/1)*
**Unit of Observation:** *Individual design (i) released by focal designer (j)*
**Model:** *Correlated Random Effects Probit Model*

| | (1) | (2) | (3) | (4) | (5) | (6) |
|---|---|---|---|---|---|---|
| | Probit Model Results | Marginal Effects | Probit Model Results | Marginal Effects | Probit Model Results | Marginal Effects |
| *Number of Followers* | 0.0500 | 0.0026 | 0.0501 | 0.0024 | 0.0398 | 0.0019 |
| | (0.0294) | (0.0016) | (0.0314) | (0.0015) | (0.0299) | (0.0014) |
| | [0.0895] | [0.0956] | [0.1109] | [0.1130] | [0.1833] | [0.1853] |
| *Derivative of Existing Design (0/1)* | | | -0.4672 | -0.0224 | -0.4855 | -0.0230 |
| | | | (0.0607) | (0.0026) | (0.0651) | (0.0027) |
| | | | [0.0000] | [0.0000] | [0.0000] | [0.0000] |
| CONTROLS | | | | | | |
| *Number of Previous Derivates* | 0.0069 | 0.0004 | | | 0.0552 | 0.0026 |
| | (0.0358) | (0.0019) | | | (0.0424) | (0.0020) |
| | [0.8483] | [0.8486] | | | [0.1932] | [0.1929] |
| *File Size* | 0.0129 | 0.0007 | 0.0145 | 0.0007 | 0.0153 | 0.0007 |
| | (0.0054) | (0.0003) | (0.0056) | (0.0003) | (0.0054) | (0.0003) |
| | [0.0176] | [0.0170] | [0.0091] | [0.0112] | [0.0049] | [0.0061] |
| *Market Tenure* | 0.0404 | 0.0021 | 0.0389 | 0.0019 | 0.0401 | 0.0019 |
| | (0.0149) | (0.0008) | (0.0153) | (0.0007) | (0.0151) | (0.0007) |
| | [0.0065] | [0.0079] | [0.0112] | [0.0113] | [0.0080] | [0.0082] |
| *Constant* | -2.0452 | | -1.9665 | | -2.0202 | |
| | (0.4185) | | (0.4099) | | (0.4155) | |
| | [0.0000] | | [0.0000] | | [0.0000] | |
| *Designer Correlated RE* | Yes | Yes | Yes | Yes | Yes | Yes |
| *Cohort FE* | Yes | Yes | Yes | Yes | Yes | Yes |
| *Category FE* | Yes | Yes | Yes | Yes | Yes | Yes |
| *Observations* | 182419 | 182419 | 182419 | 182419 | 182419 | 182419 |
| *Pseudo R²* | 0.115 | 0.115 | 0.117 | 0.117 | 0.124 | 0.124 |

*Note.* Standard errors, reported in the parentheses, are robust to heteroscedasticity and clustered at the user level. p-values are reported in brackets. Probit regression results are reported in Columns 1, 3 & 5. Corresponding Marginal effects (at means) are reported in columns 2, 4 and 6. Because the probit coefficients effects are challenging to interpret, the corresponding marginal effects provide a more intuitive interpretation. In Column 2, a one percent increase in the number of followers (log scaled variable) corresponds to a 0.026% increase in the likelihood of using a non-derivative license. As a result, a 100% increase in the number of followers (effectively a doubling of followers) would correspond to a 2.6% increase in the likelihood of using a non-commercial license. Results are not statistically significant at the 95% level in all columns (Col 1: $\beta$ = 0.0500, SE = 0.0294; p = 0.0895; Col 3: $\beta$ = 0.0502, SE = 0.0314, p = 0.1109; Col 5: $\beta$ = 0.0398, SE = 0.0299, p = 0.1833). In Models 3 − 6, the variable for whether a particular design is a derivative design is introduced. The results are negative (Col 3: $\beta$ = -.4672, SE = 0.0607; p = 0.0000; Col 5: $\beta$ = -.4855, SE = 0.0651, p = 0.0000) and significant across specifications. Results indicate that when a design is a derivative design, this would decrease the likelihood in using a non-derivative license by 2.24 percentage points (Column 4). It is important to note that the baseline rate of being a derivative is 0.15, and therefore this corresponds to an increase of 14.9% (0.0224/0.15).

**Table 7.** Results for Decision to Use No-Derivative Licenses, second-stage IV regression

**Second Stage Outcome Variable:** *Use of Non-Derivative License (0/1).*
**Instrument:** *Indicator for Period After Contributor is Features in the Community.*
**Unit of Observation:** *Individual design (i) released by focal designer (j).* **Model:** *Control Function Approach*

| | (1) | (2) | (3) |
|---|---|---|---|
| | | Second Stage | |
| Outcome Variable | | *Non-Derivative Licensee* | |
| **INSTRUMENTED VARIABLE** | | | |
| | | | |
| *Number of Followers (Instrumented)* | 0.0239 | 0.0276 | 0.0032 |
| | (0.0611) | (0.0579) | (0.0843) |
| | [0.6960] | [0.6331] | [0.9695] |
| | | | |
| **REMAINING VARIABLES** | | | |
| | | | |
| *Derivative of Existing Design (0/1)* | -0.4663 | | -0.4847 |
| | (0.0364) | | (0.0448) |
| | [0.0000] | | [0.0000] |
| *Number of Previous Derivates* | | 0.0115 | 0.0602 |
| | | (0.0311) | (0.0393) |
| | | [0.7106] | [0.1256] |
| *File Size* | 0.0153 | 0.0135 | 0.0158 |
| | (0.0043) | (0.0040) | (0.0038) |
| | [0.0004] | [0.0007] | [0.0000] |
| *Market Tenure* | 0.0454 | 0.0457 | 0.0477 |
| | (0.0155) | (0.0169) | (0.0200) |
| | [0.0034] | [0.0068] | [0.0169] |
| | | | |
| *Designer Correlated RE* | Yes | Yes | Yes |
| *Cohort FE* | Yes | Yes | Yes |
| *Category FE* | Yes | Yes | Yes |
| *Observations* | 182419 | 182419 | 182419 |
| *Pseudo R²* | 0.1200 | 0.1245 | 0.1291 |

*Note.* Standard errors, reported in the parentheses, are robust to heteroscedasticity and clustered at the user level. p-values are reported in brackets.
The first stage is the same as the baseline analyses, thus not reported here.

**Table 8.** Results for Decision to Use No-Derivative Licenses, LPM

**Outcome Variable:** *Use of Non-Derivative License (0/1)*
**Unit of Observation:** *Individual design (i) released by focal designer (j)*
**Model:** *Fixed Effects OLS (Linear Probability Model)*

| | (1) | (2) | (3) | (4) | (5) | (6) | (7) | (8) |
|---|---|---|---|---|---|---|---|---|
| *Number of Followers* | 0.0165 | 0.0058 | 0.0164 | 0.0152 | 0.0046 | 0.0046 | 0.0050 | 0.0048 |
| | (0.0003) | (0.0024) | (0.0049) | (0.0042) | (0.0024) | (0.0024) | (0.0025) | (0.0025) |
| | [0.0000] | [0.0139] | [0.0007] | [0.0003] | [0.0555] | [0.0550] | [0.0422] | [0.0502] |
| | | | | | | | | |
| *Derivative of Existing Design (0/1)* | -0.0288 | -0.0137 | -0.0282 | -0.0274 | -0.0137 | -0.0137 | | -0.0135 |
| | (0.0011) | (0.0013) | (0.0053) | (0.0039) | (0.0013) | (0.0013) | | (0.0013) |
| | [0.0000] | [0.0000] | [0.0000] | [0.0000] | [0.0000] | [0.0000] | | [0.0000] |
| CONTROLS | | | | | | | | |
| | | | | | | | | |
| *Number of Previous Derivates* | | | | | | | -0.0028 | -0.0017 |
| | | | | | | | (0.0017) | (0.0017) |
| | | | | | | | [0.0917] | [0.3100] |
| *File Size* | 0.0009 | 0.0014 | 0.0009 | 0.0002 | 0.0014 | 0.0013 | 0.0012 | 0.0013 |
| | (0.0002) | (0.0002) | (0.0024) | (0.0023) | (0.0002) | (0.0002) | (0.0002) | (0.0002) |
| | [0.0000] | [0.0000] | [0.7131] | [0.9160] | [0.0000] | [0.0000] | [0.0000] | [0.0000] |
| *Market Tenure* | -0.0008 | 0.0016 | -0.0006 | -0.0003 | 0.0010 | 0.0010 | 0.0011 | 0.0011 |
| | (0.0002) | (0.0007) | (0.0013) | (0.0013) | (0.0008) | (0.0008) | (0.0007) | (0.0007) |
| | [0.0011] | [0.0175] | [0.6549] | [0.8287] | [0.1845] | [0.1906] | [0.1257] | [0.1484] |
| *Constant* | 0.0093 | 0.0019 | 0.0089 | 0.0171 | 0.0062 | 0.0067 | 0.0068 | 0.0071 |
| | (0.0022) | (0.0035) | (0.0304) | (0.0285) | (0.0042) | (0.0043) | (0.0044) | (0.0044) |
| | [0.0000] | [0.5838] | [0.7698] | [0.5492] | [0.1415] | [0.1200] | [0.1192] | [0.1083] |
| | | | | | | | | |
| *Designer FE* | No | Yes | No | No | Yes | Yes | Yes | Yes |
| *Cohort FE* | No | No | Yes | No | Yes | Yes | Yes | Yes |
| *Category FE* | No | No | No | Yes | No | Yes | Yes | Yes |
| *Observations* | 182453 | 182453 | 182453 | 182453 | 182453 | 182453 | 182453 | 182453 |
| *Adjusted R²* | 0.0220 | 0.605 | 0.0295 | 0.0331 | 0.605 | 0.605 | 0.605 | 0.605 |

Note. Standard errors, reported in the parentheses, are robust to heteroscedasticity and clustered at the user level. p-values are reported in brackets.

**Table 9.** Results for Decision to Use No-Derivative Licenses, Fixed Effect Logit Model

**Outcome Variable:** *Use of Non-Derivative License (0/1)*
**Unit of Observation:** *Individual design (i) released by focal designer (j)*
**Model:** *Fixed Effects Logit Model*

|  | (1) | (2) | (3) |
|---|---|---|---|
| *Number of Followers* | 0.2241 | 0.2345 | 0.2251 |
|  | (0.0601) | (0.0602) | (0.0610) |
|  | [0.0002] | [0.0001] | [0.0002] |
| *Derivative of Existing* | -1.6402 |  | -1.6392 |
| *Design (0/1)* | (0.1210) |  | (0.1214) |
|  | [0.0000] |  | [0.0000] |
| CONTROLS |  |  |  |
| *Number of Previous Derivates* |  | -0.1195 | -0.0085 |
|  |  | (0.0817) | (0.0853) |
|  |  | [0.1438] | [0.9202] |
| *File Size* | 0.1027 | 0.0950 | 0.1027 |
|  | (0.0132) | (0.0131) | (0.0132) |
|  | [0.0000] | [0.0000] | [0.0000] |
| *Market Tenure* | 0.1299 | 0.1316 | 0.1303 |
|  | (0.0357) | (0.0355) | (0.0359) |
|  | [0.0003] | [0.0002] | [0.0003] |
| *Designer* FE | Yes | Yes | Yes |
| *Cohort* FE | Yes | Yes | Yes |
| *Category* FE | Yes | Yes | Yes |
| *Observations* | 16787 | 16787 | 16787 |
| *Pseudo* $R^2$ | 0.0624 | 0.0375 | 0.0624 |

Note. Standard errors, reported in the parentheses, are robust to heteroscedasticity and clustered at the user level. p-values are reported in brackets.

**Table 10.** The effect on Closedness of Licenses

**Outcome Variable:** *Closedness (0/1/2/3/4)*
**Unit of Observation:** *Individual design (i) released by focal designer (j)*
**Model:** *Ordered Probit (column (1)), OLS (column (2))*

|  | (1) | (2) |
|---|---|---|
| *Number of Followers* | 0.0507 | 0.0302 |
|  | (0.0176) | (0.0108) |
|  | [0.0041] | [0.0053] |
|  |  |  |
| *Derivative of Existing Design (0/1)* | -0.3572 | -0.2176 |
|  | (0.0145) | (0.0079) |
|  | [0.0000] | [0.0000] |
| CONTROLS |  |  |
|  |  |  |
| *Number of Previous Derivates* | 0.0418 | 0.0263 |
|  | (0.0139) | (0.0084) |
|  | [0.0026] | [0.0018] |
| *File Size* | 0.0133 | 0.0089 |
|  | (0.0020) | (0.0009) |
|  | [0.0000] | [0.0000] |
| *Market Tenure* | -0.0266 | -0.0147 |
|  | (0.0055) | (0.0034) |
|  | [0.0000] | [0.0000] |
| *Constant* |  | 1.3983 |
|  |  | (0.0316) |
|  |  | [0.0000] |
|  |  |  |
| *Designer FE* | Yes | Yes |
| *Cohort FE* | Yes | Yes |
| *Category FE* | Yes | Yes |
| *Observations* | 182453 | 182453 |
| *Adjusted* $R^2$ |  | 0.6173 |
| *Pseudo* $R^2$ | 0.0827 |  |

Note. Standard errors, reported in the parentheses, are robust to heteroscedasticity and clustered at the user level. p-values are reported in brackets.

# APPENDIX

## A.1 Additional Information on Licenses

Approximately 98% of the designs in our empirical setting are licensed under *Creative Commons* licenses, which are based on combinations of the following modules that each specify certain aspects of how the licensed product can or cannot be used:

- **Attribution:** Imposes a constraint that follow-on users must acknowledge or cite the original source.

- **Non-Commercial:** Indicates that follow-on users cannot use both the product and its derivative products for commercial purposes.

- **Share Alike:** Indicates that the follow-on product must be shared under the same terms as the product being reused.

- **No Derivative:** Indicates that this product cannot be used to create derivative products.

These modules, in turn, can be combined into different licenses:

- **Attribution:** This license only contains the "Attribution" module.

- **Attribution – Share Alike:** This license allows commercial use and reuse (i.e., the creation of derivatives), requires attribution, and mandates that any derivatives be shared under the same license.

- **Attribution – No Derivative:** This license allows for commercial use and requires attribution but does not allow for the creation of derivatives.

- **Attribution – Non-Commercial:** This license allows for derivatives and requires attribution, but it does not allow for commercial use. While new products based on another creator's product must acknowledge the creator and be non-commercial, the creators of the derivative are not obliged to license their product (i.e., the derivative) under the same license. This means that if someone modifies a creator's work, they are not required to distribute the modified work under the same license. They are free to choose any license for their modifications as long as they are not using it commercially.

- **Attribution – Non-Commercial – Share Alike:** This license permits derivatives and requires acknowledgement of the original creator. It also mandates that any derivative must be shared under the same license. However, it does not allow for commercial use. With this license, others are allowed to tweak, and build upon (i.e., reuse) a creator's work for non-commercial purposes, as long as they give credit to that creator and license their new creations under identical terms. Therefore, if someone modifies a creator's work, they are obliged to distribute the modified work under the same "Attribution – Non-Commercial – Share Alike" license.

- **Attribution – Non-Commercial – No Derivative:** This license requires acknowledgement of the original creator when used but does not allow for commercial use or the creation of derivatives.

In addition to these licenses that result from the combination of different modules, individuals can also choose among a set of other licenses:

- **All Rights Reserved:** This is the most restrictive license, granting no rights for commercial use or the creation of derivatives.

- **Public Domain Dedication:** This license indicates that the product is released into the public domain, meaning that the product is not protected by copyright and can be used freely by anyone without obtaining a license or providing attribution.

- **GNU - GPL:** This license permits commercial use and the creation of derivatives, and also requires sharing alike. However, it does not require attribution or release of the reused code.

- **GNU Lesser General:** This license permits commercial use and the creation of derivatives, but it does not require attribution or sharing alike. However, any modifications to the licensed product itself must be released under the same license.

- **BSD License:** This license permits commercial use and the creation of derivatives, and also releases the work to the public domain. It does not require attribution or sharing alike.

- **Nokia:** This license permits the creation of derivative works and requires attribution and sharing alike, but it does not allow for commercial use or release to the public domain.

*Table A1* provides an overview of the licenses that are used in our empirical setting and how they relate to our outcome variables of interest.

--------------------------------------------
Insert *Table A1* about here
--------------------------------------------

## A.2 Sensitivity of Instrumental Variable Analysis

*Table A2* shows a sensitivity analysis grounded in the methodology outlined by Cinelli & Hazlett (2020). This analysis enables us to gauge the effect of unobserved variables on the coefficients. The outcomes were obtained using the *sensemakr* package in *STATA*. $R^2_{Y \sim Z|X,D}$ represents the extent to which an extreme confounder, one that accounts for 100% of the variance, would need to explain the remaining variance in the treatment in order to completely explain away the observed effect. In simpler terms, $R^2_{Y \sim Z|X,D}$ measures the degree to which a hypothetical maximum confounder must influence the unexplained variance in the treatment to entirely negate the observed impact. $RV_{q=1}$ signifies the portion of the residual variance, in both the treatment and the outcome, that must be explained to completely eliminate the observed effect, essentially reducing the coefficient to zero. $RV_{q=1, \alpha=0.05}$ indicates the threshold value necessary to shift the coefficient into a range where it loses its statistical significance. In other words, the treatment's impact becomes statistically indistinguishable from zero, suggesting that it does not have a significant effect on the outcome.

One challenge with interpreting the aforementioned values lies in their lack of intuitiveness. For instance, unobserved confounders (orthogonal to the covariates) that explain more than 13.14% of the residual variance in both the treatment and the outcome have enough strength to reduce the point estimate to 0, effectively causing a bias that amounts to 100% of the original estimate. On the other hand, confounders that account for less than 12.74% of the residual variance in both the treatment and the outcome lack the strength to drop the point estimate to 0.

To convey these effects more intuitively, we compare them to the scale of observed confounding variables. We selected the *File Size* variable for this purpose, given its strong correlation with both the outcome variable and *Number of Followers*. The results illustrate potential changes to the coefficients of *Number of Followers* if the residual variance was 10 times, 50 times, or even 100 times the explanatory power of *File Size*.

For example, we demonstrate that if an unobserved variable has 10 times the magnitude of *File Size*, the coefficient in an OLS regression with category and cohort fixed effects, originally 0.0461 (see *Table A2*), would drop to 0.0398 (t = 1.863). However, to drive the coefficient to zero, the unobserved variance should be more than 50 times the observed variance. Given that *File Size* is a crucial covariate explaining a substantial portion of the variation in both the outcome variable and *Number of Followers* (as shown in the correlation table), this strongly suggests that unobserved effects would need to be significantly large to nullify the patterns we have observed.

## A.3 Additional Figures and Tables

**Table A1.** Explanations of different licenses

| License Type | License | Non-Commercial License (0/1) | Non-Derivative License (0/1) | Closedness (0/1/2/3/4) |
|---|---|---|---|---|
| | Public Domain Dedication | 0 | 0 | 0 |
| | Attribution | 0 | 0 | 1 |
| | Attribution – Share Alike | 0 | 0 | 2 |
| Creative | Attribution – Non-Commercial | 1 | 0 | 2 |
| Commons | Attribution – No Derivative | 0 | 1 | 3 |
| | Attribution – Non-Commercial – Share Alike | 1 | 0 | 3 |
| | Attribution – Non-Commercial – No Derivative | 1 | 1 | 4 |
| | All Rights Reserved | 1 | 1 | 4 |
| GNU | GNU - GPL | 0 | 0 | 0 |
| GNU | GNU Lesser General | 0 | 0 | 1 |
| BSD | BSD License | 0 | 0 | 1 |
| Nokia | Nokia | 1 | 0 | 3 |

Note. The "No Derivative" module implies that the product cannot be modified or used to create derivatives. Therefore, the "Share Alike" module is not relevant when users specify the "No Derivative" module.

**Table A2.** Sensitivity of Main Results to Unobserved Confounders

| Variable | Coef. | S.E. | t-value | $R^2_{Y \sim Z|X,D}$ | $RV_{q=1}$ | $RV_{q=1, \alpha=0.05}$ |
|---|---|---|---|---|---|---|
| *Number of Followers* | 0.0461 | 0.0008 | 60.2033 | 0.0195 | 0.1314 | 0.1274 |

| Illustrative Bound | Coef. | S.E. | t-value | $R^2_{Y \sim Z|X,D}$ | $RV_{q=1}$ | 95% CI [Lower, Upper] |
|---|---|---|---|---|---|---|
| 10.00 × *File Size* | 0.0398 | 0.0008 | 50.721 | 0.0071 | 0.0506 | [0.0382, 0.0413] |
| 50.00 × *File Size* | 0.0101 | 0.0009 | 11.583 | 0.0358 | 0.2530 | [0.0084, 0.0118] |
| 100.00 × *File Size* | -0.0428 | 0.0011 | -40.743 | 0.0721 | 0.5061 | [-0.0449, -0.0408] |

**Table A3.** IV regression with a matched sample

**First Stage Outcome Variable:** *Number of Followers.*
**Second Stage Outcome Variable:** *Use of Non-Commercial License (0/1).*
**Instrument:** *Indicator for Period After Contributor is Features in the Community.*
**Unit of Observation:** *Individual design (i) released by focal designer (j).*
**Model:** *Control Function Approach*

|  | (1) | (2) |
|---|---|---|
| **INSTRUMENTED VARIABLE** | | |
| | | |
| *Number of Followers (Instrumented)* | | 0.5875 |
| | | (0.1053) |
| | | [0.0000] |
| *Featured (Instrument)* | 1.1675 | |
| | (0.2700) | |
| | [0.0000] | |
| | | |
| **REMAINING VARIABLES** | | |
| | | |
| *Derivative of Existing Design (0/1)* | -0.1601 | 0.1714 |
| | (0.0677) | (0.1243) |
| | [0.0194] | [0.1681] |
| *Number of Previous Derivates* | 0.4598 | -0.2623 |
| | (0.2580) | (0.1187) |
| | [0.0768] | [0.0271] |
| *File Size* | -0.0270 | 0.0128 |
| | (0.0118) | (0.0186) |
| | [0.0240] | [0.4911] |
| *Market Tenure* | 0.5043 | -0.2725 |
| | (0.1057) | (0.0714) |
| | [0.0000] | [0.0001] |
| | | |
| *Designer FE* | Yes | Yes |
| *Cohort FE* | Yes | Yes |
| *Category FE* | Yes | Yes |
| *Observations* | 4785 | 72431 |
| *Adjusted* $R^2$ | 0.671 | 0.0259 |

*Note.* Standard errors, reported in the parentheses, are robust to heteroscedasticity and clustered at the user level. p-values are reported in brackets. Instrument and first stage model validity: Given that the reduced form for the endogenous explanatory variable is linear, we use the same diagnostics as in the linear case. The Cragg–Donald Wald F-test statistic rejects the null hypothesis of a weak instrument. The test statistic is 779.764, which exceeds the critical value of 16.38 proposed by Stock and Yogo (2005). The Kleibergen–Paap rk Wald F-statistics is 675.966, which also allays concerns over a weak instrument